\documentclass{article}
	\usepackage{fullpage, graphicx,setspace}
	\author{Sujit S. Datta\thanks{E-mail: \texttt{ssdatta@physics.harvard.edu}}\\ {\normalsize Department of Physics, Harvard University, Cambridge MA 02138}}	
	\date{{\normalsize \today}}
	\title{Wetting and Energetics in Nanoparticle Etching of Graphene}
	\begin{document}
	\doublespacing

	\maketitle
	\begin{abstract}
Molten metallic nanoparticles have recently been used to construct graphene nanostructures with crystallographic edges. The mechanism by which this happens, however, remains unclear. Here, we present a simple model that explains how a droplet can etch graphene. Two factors possibly contribute to this process: a difference between the equilibrium wettability of graphene and the substrate that supports it, or the large surface energy associated with the graphene edge. We calculate the etching velocities due to either of these factors and make testable predictions for evaluating the significance of each in graphene etching. This model is general and can be applied to other materials systems as well. As an example, we show how our model can be used to extend a current theory of droplet motion on binary semiconductor surfaces.
\end{abstract}

\begin{center}
\line(1,0){300}\\ 
\end{center}
		\doublespacing

\section{Introduction}
Graphene-derived nanomaterials are emerging as a unique class of low-dimensional structures, in part due to their remarkable electronic properties and intriguing chemical reactivity \cite{geim1,geim2}. Graphene's two-dimensionality makes it compatible with current planar device architectures. In addition, its electronic properties can be a strong function of how it is structured \cite{nr1,nr2,nr3,nr4,nr5}. Because of these characteristics, a significant amount of research is currently focused on constructing graphene-based nanoelectronic devices, which hold enormous promise for high-performance post-silicon applications. Desirable devices include nanoribbons with crystallographic edges, which would potentially combine the exceptional electronic properties of graphene with rational control of device characteristics by modification of the graphene architecture.

Experimentally, constructing these devices is a formidable challenge \cite{geim2}. Most current nanolithography techniques produce graphene structures with rough, noncrystalline edges, which can be quite detrimental to device performance \cite{geim2,nr1,nr6,nr7,nr8,nr9,nr10}. Alternative methods are being explored, although a good deal of work is required to optimize and extend them to practical usage. In previous work, my colleagues and I developed a technique by which molten metallic nanoparticles etch straight trenches in few-layer graphene flakes supported on amorphous SiO$_{2}$ \cite{datta1}. Strikingly, we found that these trenches span lengths of $>1\mu$m along directions commensurate with the graphene honeycomb lattice and extend down to the underlying substrate through the graphene. Further work has shown the edges of these trenches to be crystallographic \cite{previous1}.

The microscopic mechanism by which this crystallographic etching occurs is unclear. Here, we propose a simple model for this process, motivated by the results of etching experiments as well as theoretical models for ``running" droplets \cite{degennes1,ondarcuhu}. Two factors potentially drive crystallographic etching by moving droplets: a difference between the equilibrium wettability of the graphene and the underlying substrate, or the high surface energy associated with a reactive graphene edge. The analysis presented here yields a number of testable predictions that could be used to characterize the significance of each of these factors in controlling graphene etching. Furthermore, the ideas presented here are relevant to a number of other materials systems as well. As an example, we study the case of droplets formed during the evaporation of binary semiconductors. We use our model to show how a current theory of this process may be extended to better agree with experimental observations.

\section{Model and Results}
\subsection{Experimental Motivation}
As reported previously \cite{datta1}, a commonly-used preparation process for etching of few-layer graphene samples supported on SiO$_{2}$ involves uniformly coating them with a solution of iron salt. These samples are then heated in a hydrogen/argon environment at 900$^\circ$C for a given period of time, typically on the order of 1 hour. Under these conditions, the iron salt forms nanoparticles with diameter $\sim20$nm \cite{klinke}. Because this temperature is above the melting point of the nanoparticles, they form molten droplets \cite{molten1,molten2}. Driven by Brownian motion, these diffuse over the surface of the amorphous SiO$_{2}$ substrate or the graphene before encountering a graphene edge. 

Because they are not part of the bulk $sp^{2}$-bonded network, carbon atoms at graphene edges are chemically reactive \cite{reactive1,reactive2,reactive3}. (On the other hand, atoms {\it not} at edges are inert and likely do not react with the molten droplets. This is supported by observations that etched trenches begin only at graphene edges.) These atoms likely undergo catalytic hydrogenation, a chemical reaction catalyzed by the molten Fe droplet in which the carbon atoms react with the hydrogen supplied from the surrounding environment, forming methane gas \cite{datta1}. This process is thus initiated when the thermally-driven droplet randomly encounters such an edge. The droplet removes the contacting atoms from the graphene sheet, forming a trench, and propagates the reactive edge further into the graphene sheet -- we term this the {\it step edge}. This is shown schematically in Figure 1. Experimental observations show these trenches to be straight over micrometers, thus indicating that lateral motion of the droplet due to thermal effects can be neglected while considering the etching process. Thermal effects may become significant for droplets much smaller than those considered here. 

This procedure has since been extended to the etching of monolayer graphene, the surface of bulk highly-oriented graphite, and carbon nanotubes using a variety of metal catalyst materials and at a range of high temperatures \cite{jarillo,previous1,previous2,previous3,previous4,previous5,previous6, previous7}. The temperatures used in these experiments are similar to or higher than the nanoparticle melting temperature, which is well known to be significantly lower than the melting temperature of the bulk material \cite{meltingdepression}. Thus, the nanoparticles are likely to either be fully molten or have a molten surface layer. Furthermore, the etched trenches and the manner in which they form appear to be similar in these experiments. Taken together, these observations indicate that metal nanoparticle etching of graphene at high temperatures may have a common origin, and a model treating the nanoparticles as chemically-reactive droplets may capture the underlying physics of this process. Two criteria guide the continued etching along straight crystallographic lines: the choice of the initial direction of the droplet and the constant velocity of the droplet along that direction. 

\subsection{Outline Of The Model}
The choice of the initial direction of the droplet as it meets the graphene edge is a result of the chemistry of the etching process. Experimental observations indicate that the etched trenches extend down to the underlying substrate \cite{datta1}, suggesting that removal of carbon is fast compared to the motion of the droplet. Experiments on bulk samples \cite{anisotropic1} as well as density-functional theory simulations \cite{previous1} indicate that this process is likely to be strongly anisotropic. That is, the energy required for removal of carbon atoms from a graphene sheet by catalytic hydrogenation is significantly reduced along particular crystallographic directions. Thus, when a molten droplet meets a reactive graphene edge, etching occurs fastest along whichever crystallographic direction requires the least amount of energy for carbon removal to occur. While this step is not the focus of this paper, we stress that a deeper understanding of it is crucial to a full understanding of nanoparticle etching of graphene. 

Once the carbon atoms in contact with the molten droplet are removed and the initial droplet direction is chosen, the droplet moves along the crystallographic direction. The droplet must continue along that direction for successful etching to occur -- this step is the main focus of this paper. We model the molten nanoparticle as a liquid droplet of characteristic size $r$, density $\rho$, and liquid-vapor surface tension $\gamma_{lv}$. For the iron nanoparticles used in graphene etching, $r\approx20$nm \cite{datta1}. Published values of the density and surface tension give $\rho\approx7$g/cm$^{3}$ and $\gamma_{lv}\approx2$N/m \cite{density,angle2}. These values yield a capillary length $\lambda=\sqrt{\gamma_{lv}/\rho g}\approx$ 5mm, where $g$ is gravitational acceleration -- because $r/\lambda\ll1$, we can safely assume that gravitational effects are negligible. Experimental etching velocities are on the order of nanometers per second \cite{datta1}, implying that  the particle Reynolds number Re $=\rho_{0}Vr/\eta_{0}\ll1$, where $\rho_{0}$ and $\eta_{0}$ are the density and viscosity of air and $V$ is the etching velocity. The minimal model presented here focuses on key details of graphene etching by nanoparticles to obtain a qualitative understanding of this process. Specifically, we derive the existence of a force driving droplet motion in one direction and investigate the dependence of the droplet velocity on relevant physical parameters. Potential extensions of this model could explicitly treat hydrodynamic flows within the droplet, the atomistic details of the graphene edge, and the effects of non-idealities such as contact-line pinning and hysteretic effects resulting from surface heterogeneities. Furthermore, one could incorporate electrostatic effects that may influence droplet characteristics \cite{mele}.

At a given time after the choice of the initial etching direction, the droplet straddles the step edge. For simplicity, we consider a two-dimensional cross-section, as shown in Figure 1B. The advancing side is on the graphene at point $g$ with advancing contact angle $\theta_{g}^{a}$, while the receding side of the droplet lies on the substrate at point $s$ with receding contact angle $\theta_{s}^{r}$. The equilibrium contact angles of the droplet deposited {\it fully} on graphene or on the substrate are defined to be $\theta_{g}$ and $\theta_{s}$, respectively. The graphene-liquid, graphene-vapor, substrate-liquid, and substrate-vapor interfacial energies are $\gamma_{gl}, \gamma_{gv}, \gamma_{sl}$, and $\gamma_{sv}$, respectively. The thermodynamic driving force for droplet motion is a result of the unbalanced Young's force at the contact line, and so the total driving force on the droplet goes as
\begin{eqnarray}
F_{\theta}\sim r\cdot[(\gamma_{gv}-\gamma_{gl})-(\gamma_{sv}-\gamma_{sl})]\\=r\cdot\gamma_{lv}(\mbox{cos}~\theta_{g}^{a}-\mbox{cos}~\theta_{s}^{r})
\end{eqnarray}
There are thus two factors that will drive a droplet to keep moving in one direction: (1) a difference between the equilibrium wettability of graphene and of the substrate that supports it, in which case $\theta_{g}^{a}$ and $\theta_{s}^{r}$ are determined primarily by $\theta_{g}$ and $\theta_{s}$, or (2) the large surface energy associated with the graphene step edge, in which case $\theta_{g}^{a}$ and $\theta_{s}^{r}$ are determined primarily by the magnitude of the energy cost associated with having a step edge. We first focus on case (1) and ignore the structure of the graphene step edge.

\subsection{Motion Driven By Differing Wettabilities }
As the droplet straddles the graphene step edge, it will remove the underlying reactive carbon via catalytic etching. To relate the thermodynamic driving force to the nature of the underlying substrate, we adapt a model proposed by Brochard and de Gennes \cite{degennes1}. For the given droplet configuration, we assume an areal density of carbon atoms that monotonically varies from 1 at the droplet advancing side to a value $\phi<1$ at the droplet receding side. In general, we expect the surface energies $\gamma_{sl}, \gamma_{sv}, \gamma_{gl},$ and $\gamma_{gv}$ to depend on $\phi$. Specifically, we expect the difference between $(\gamma_{sl}-\gamma_{sv})$ and $(\gamma_{gl}-\gamma_{gv})$ to increase with decreasing $\phi$. To first order in $\phi$, this is manifested in the relation $(\gamma_{sl}-\gamma_{sv})-(\gamma_{gl}-\gamma_{gv})=\gamma_{0}(1-\phi)$, where $\gamma_{0}$ is a parameter that reflects the increase in the differential surface energy for a reaction that has achieved completion. Relating this to the equilibrium contact angles via the Young equations $\gamma_{gv}-\gamma_{gl}=\gamma_{lv}\mbox{cos}~\theta_{g}$ and $\gamma_{sv}-\gamma_{sl}=\gamma_{lv}\mbox{cos}~\theta_{s}$, we obtain 
\begin{eqnarray}
\gamma_{lv}(\mbox{cos}~\theta_{g}-\mbox{cos}~\theta_{s})=\gamma_{0}(1-\phi)
\end{eqnarray}
Assuming that the equilibrium wettabilities of the graphene and of the substrate are the prime determinant of $\theta_{g}^{a}$ and $\theta_{s}^{r}$, as in case (1), we consider the limit where $\theta_{g}^{a}\approx\theta_{g}$ and $\theta_{s}^{r}\approx\theta_{s}$. As a result, $F_{\theta}/r\approx\gamma_{0}(1-\phi)$. The chemistry of the etching process determines $\phi$. Assuming first-order reaction kinetics, $\phi= e^{-r/V\tau}$, where $V$ is the velocity of the droplet and $\tau$ is the characteristic time constant of the reaction. This yields a direct relationship between the dynamics of the etching reaction, quantified by the time constant $\tau$, and the thermodynamic driving force for droplet motion, $F_{\theta}$:
\begin{eqnarray}
F_{\theta}\sim r\cdot\gamma_{0}(1-e^{-r/V\tau})
\end{eqnarray}

We treat the time-averaged droplet motion as a damped response to the force driving droplet motion, balancing $F_{\theta}$ with a drag force $F_{\xi}=\xi V$, where $\xi$ is the drag coefficient. Equating these forces results in a transcendental equation for the droplet velocity:
\begin{eqnarray}
V\sim\alpha(1-e^{-\beta/V});~~\alpha\equiv\gamma_{0}r/\xi,~\beta\equiv r/\tau
\end{eqnarray}
Crucially, the droplet velocity depends on two characteristic velocities, $\alpha$ and $\beta$, or what we term the {\it driving velocity} and the {\it reaction velocity}, respectively. The numerical solution to Equation 4 is presented in Figure 2. The droplet velocity increases monotonically with increasing $\alpha$ or increasing $\beta$.

We make two assumptions to directly relate $\alpha$ to experimentally measurable quantities. First, $\theta_{g}^{a}\rightarrow\theta_{g}$ and $\theta_{s}^{r}\rightarrow\theta_{s}$ in the limiting case of $\phi\rightarrow0$ -- Equation 2 then gives $\gamma_{0}=\gamma_{lv}(\mbox{cos}~\theta_{g}-\mbox{cos}~\theta_{s})$. Second, we explicitly consider the nature of the drag coefficient $\xi$ in Equation 4. This can generally be written as $\xi=\eta r f(r,\theta^{*})$, where the dynamic contact angle $\theta^{*}\equiv\mbox{cos}^{-1}\left[(\mbox{cos}~\theta_{g}^{a}+\mbox{cos}~\theta_{s}^{r})/2\right]$, $f$ is a function of $r$ and $\theta^{*}$ appropriate to the system under consideration, and $\eta$ is the droplet viscosity. Because we assume in this section that equilibrium wettabilities are the strongest determinant of the advancing and receding contact angles, we take $\theta^{*}\approx\mbox{cos}^{-1}\left[(\mbox{cos}~\theta_{g}+\mbox{cos}~\theta_{s})/2\right]$. $f(r,\theta^{*})$ now remains to be determined. 

Following previous work \cite{dissipation1}, we assume that dissipation occurs primarily via viscous processes at the droplet-substrate contact line, as is typically assumed when considering low Reynolds number free surface flows with a moving contact angle. In particular, we apply the lubrication approximation to a ``wedge" of angle $\theta^{*}$, assuming a two-dimensional Poiseuille velocity field in the wedge \cite{dissipation2}. This yields $f(r,\theta^{*})\propto l/\theta^{*}$, where the geometrical details of the system are lumped into $l\equiv\mbox{ln}|x_{max}/x_{min}|$, a logarithmic function of two parameters that describe the geometry of the region within which dissipation is concentrated. $x_{max}$ is typically on the order of the droplet size ($\approx10-100$nm), while $x_{min}$ is typically taken to be on the scale of individual molecules ($\approx0.1-1$nm), giving $l\sim1-10$. Combining these results gives $\xi\propto\eta r/\mbox{cos}^{-1}\left[(\mbox{cos}~\theta_{g}+\mbox{cos}~\theta_{s})/2\right]$. 

We need to consider the validity of the lubrication approximation, which strictly speaking only applies to small values of $\theta^{*}$, in the situation presented here for a broad range of contact angles $\leq180^\circ$. For example, dissipation may cease to be localized at the droplet-substrate contact line for sufficiently large values of the contact angle. However, because of the large viscosity of the droplet relative to the medium surrounding it, it is unlikely that the scaling behavior will change in the situation considered here. Indeed, the more general case of larger $\theta^{*}$ values has been considered theoretically. While the $\sim1/\theta^{*}$ scaling of the dissipation presented here crosses over to a $\sim\theta^{*2}$ scaling with increasing contact angle, it is important to note that this crossover occurs when $\theta^{*}=(\eta_{0}/\eta)^{-1/3}$ \cite{dissipation3}. Using Sutherland's empirical formula \cite{sutherland} to estimate the viscosity of the air surrounding the droplet gives $\eta_{0}\approx0.02-0.06$mPa$\cdot$s, while literature values of liquid metal viscosities are in the range $\eta\sim2-6$mPa$\cdot$s. These parameters yield a crossover contact angle $>180^\circ$, indicating that the $\sim1/\theta^{*}$ scaling of dissipation considered here is appropriate for our system. We thus have the characteristic velocities explicitly in terms of the experimentally-measurable equilibrium contact angles:
\begin{eqnarray}
\alpha\propto\gamma_{lv}\eta^{-1}(\mbox{cos}~\theta_{g}-\mbox{cos}~\theta_{s})\cdot\mbox{cos}^{-1}\left[(\mbox{cos}~\theta_{g}+\mbox{cos}~\theta_{s})/2\right],~\beta=r/\tau
\end{eqnarray}

These results have important implications. As shown in Figure 3, the sign and magnitude of the droplet velocity is strongly dependent on the values of $\theta_{g}$ and $\theta_{s}$. Notably, $V>0$ only when $\theta_{g}<\theta_{s}$. While only a few clear equilibrium contact angle measurements of $\theta_{g}$ and $\theta_{s}$ exist for typical nanoparticle materials, these data suggest that the mechanism driving droplet motion outlined in case (1) may play a significant role in graphene etching, as shown in Figure 3. More details are provided in Appendix A. 

\subsection{Motion Driven By Edge Energetics}
The structure of the graphene step edge may also contribute to driving droplet motion in graphene etching. This driving force is the dominant mechanism of case (2), in which the equilibrium wettabilities of graphene or the substrate do not play a significant role. Instead, the difference between the surface energies $\gamma_{gv}$ and $\gamma_{sv}$ (and hence, the values of $\theta_{g}^{a}$ and $\theta_{s}^{r}$) will be determined primarily by the energetics of the step edge. We will consider this case using a treatment similar to that used in Section 2.3.

Because the atoms at a freshly-created step edge have dangling bonds, we associate a surface energy with each step \cite{edgeenergy1,edgeenergy2,edgeenergy3,edgeenergy4}. We make the simplifying assumption that $\gamma_{gl}=\gamma_{sl}$ and $\gamma_{gv}=\Delta+\gamma_{sv}$, where $\Delta$ is the total surface energy density associated with the overall step structure across the droplet profile. This step structure is shown schematically in Figure 1. Thus, as given by Equation 1, the driving force for droplet motion goes as $F_{\theta}\sim r\cdot\Delta$. We adopt the previous discussion to this situation: coarse-graining over the atomic details of the step edge, we assume an areal density of steps that monotonically varies from 1 at the droplet advancing side to a value $\phi<1$ at the droplet receding side, such that $\Delta=\Delta_{0}(1-\phi)$. The results of the previous case thus carry over, with the key replacement of the phenomenological surface energy $\gamma_{0}$ with the step energy $\Delta_{0}$. The droplet velocity again depends on the two characteristic velocities $\alpha$ and $\beta$ as given by Equation 4. As in case (1), the droplet velocity increases monotonically with increasing $\alpha$ or increasing $\beta$, as shown in Figure 2. However, $\alpha$ is now determined by different physical quantities as manifested in the new relation $\alpha\equiv\Delta_{0}r/\xi$. 

We can relate the step energy to the structure of the step \cite{edgeenergy1,edgeenergy2,edgeenergy3,edgeenergy4,edgeenergy5}. In particular, assuming that the energy of an individual step is proportional to the number of atoms making up the step, $\Delta_{0}\propto(t/a)\cdot E_{bond}$ where $t$ is the step height, $a$ is the average bond length of the graphene in the direction normal to its surface, and $E_{bond}$ is the energy required to form one dangling bond, with the proportionality constant incorporating the geometric details of the step structure. 

Unlike the previous case, we cannot express the advancing and receding contact angles $\theta_{g}^{a}$ and $\theta_{s}^{r}$, and thus the drag coefficient $\xi$, in terms of experimentally-determinable equilibrium contact angles. However, as in the previous case, we can assume that dissipation occurs primarily at the droplet-substrate contact line. This gives $\xi\propto\eta r$, and hence
\begin{eqnarray}
\alpha\propto(t/a)\cdot E_{bond}~\eta^{-1},~\beta=r/\tau
\end{eqnarray}

Finally, we note that because the catalytic hydrogenation reaction occurs very quickly, as noted in Section 2.2, the reaction time constant $\tau$ is likely to be small relative to the time the droplet takes to move its own length. It is thus probable that the etching process is not limited by the speed of the reaction and $\alpha$ is the key determinant of the droplet velocity for both of the driving forces described here. Furthermore, this supports our assumption that the metal catalyst is not saturated with an appreciable amount of carbon during the etching process and does not lose catalytic behavior as the reaction progresses.

\section{Discussion}
\subsection{Relevance To Graphene Etching}
We have focused our discussion on developing predictions for the time-averaged etching velocity, $V$. Not only is maximizing $V$ crucial to optimizing this etching process, but it is also likely to be the most easily measurable physical parameter in experiments. Measuring the etching velocity can be done in a number of ways. One possibility is to use real-time transmission electron microscopy \cite{realtime}, but this can be experimentally challenging. Another approach is to pattern the catalyst particles at a graphene step edge \cite{previous5}. Measuring the length of the trenches formed after a given etching time directly yields the distribution of time-averaged etching velocities. 

The force driving droplet motion in graphene etching is likely to have contributions from both factors described in Sections 2.3 and 2.4 -- namely, (1) a difference between the equilibrium wettability of graphene and of the substrate that supports it and (2) the high surface energy associated with forming a graphene step edge. An outstanding question is whether one of these is a stronger determinant of droplet motion than the other. Crucially, different physical parameters determine the contributions from factors (1) and (2), enabling their importance to be tested experimentally. 

If factor (1) plays a significant role, etching velocities will be a strong function of the wettability of the underlying substrate, as indicated in Equation 5. This effect can be tuned by studying etching of the same material on entirely different substrates. To assess the experimental feasibility of this test of factor (1), we list in Table 1 literature values of $\theta_{s}$ for Fe, Ni, Co, and Ag droplets on a number of materials that are commonly used as substrates for graphene \cite{al2o3, si3n4, deheer}. Comparing these to the data presented in Figure 3 suggests that if factor (1) influences the motion of droplets, the etching velocities for Fe should follow the trend $v_{SiO_{2}}>v_{Al_{2}O_{3}}>v_{Si_{3}N_{4}}>v_{SiC}$, where $v_{i}$ indicates the etching velocity on substrate $i$. Furthermore, the etching velocities of Ni or Co on SiO$_{2}$, Al$_{2}$O$_{3}$ and Si$_{3}$N$_{4}$ should be similar to each other, and significantly greater than the etching velocities of Ni or Co on SiC. The case of Ag is a unique one. Because we do not expect factor (1) to contribute to its etching behavior, the etching velocities of Ag should be independent of the substrate used. A more practical alternative is to study etching on a substrate that has been surface-treated to various degrees, for example, by coating it with different concentrations of a suitable chemical species. The etching velocities will be a strong function of the wettability of the graphene itself, as well. The graphene wettability can be varied via suitable surface treatment, such as by using a range of strengths of oxygen plasma etching \cite{surfaceenergy1} or by sonicating the graphene in different solvents \cite{surfaceenergy2}. 

If factor (2) plays a significant role, etching velocities will be a strong function of the thickness of the graphene. A systematic study of etching velocity as a function of graphene thickness remains to be done. The thickness is easily varied experimentally. For example, micromechanical cleavage of highly-oriented pyrolytic graphite using Scotch tape can result in monolayer graphene flakes and few-layer graphene flakes with thicknesses up to tens or hundreds of layers. As described in Section 2.4, the etching velocity should increase linearly with the thickness of the graphene. Performing this study using Ag droplets would be particularly interesting. Because we have ruled out the possible contribution of factor (1) to etching by Ag, as described in Appendix A and Figure 3, we expect factor (2) to be the only driving force for etching. 

\subsection{Relevance To Binary Semiconductors}
``Running" droplets similar to those considered here form on the surface of evaporating binary semiconductors. Unlike the inert substrate in the case of graphene etching, the underlying semiconductor substrate acts as a reservoir for droplet formation. Experimental advances have led to the direct imaging of droplet formation and movement. As recently shown by Tersoff {\it et al.} \cite{tersoff},  droplets form on the surface of the binary compound when it is annealed at high enough temperature, move in directions commensurate with the underlying crystal lattice, and leave behind a trail potentially similar to the etched trenches described earlier for graphene etching \cite{steps}. To explain this, they model this process by considering the thermodynamics of noncongruent evaporation, in which one chemical component of the binary compound evaporates slower than the other. We believe that this model is incomplete and propose that the ideas presented in Section 2.4 can provide a more complete physical picture. Here we briefly summarize Tersoff {\it et al.}'s argument and extend it using our model.

For temperatures $T$ higher than $T_{c}$, the upper limit for congruent evaporation, Ga forms droplets on the GaAs crystal surface because the Ga chemical potential of the surface $\mu_{s}$ is larger than the liquidus chemical potential $\mu_{l}$. If one assumes the GaAs crystal surface to be totally isotropic, the driving force for droplet motion presented in Equation 1 is zero, since $\gamma_{gv}=\gamma_{sv}$ and $\gamma_{gl}=\gamma_{sl}$. Tersoff {\it et al.}'s model breaks this symmetry by assuming that while the GaAs surface is topographically unchanged, a droplet displacement to one side (e.g. by thermal fluctuations) will cause the surface energy of the newly-exposed part of the GaAs surface to be at $\gamma_{sv}=\gamma_{sv(l)}$. On the other hand, the newly-covered part of the GaAs surface will still remain at $\gamma_{gv}=\gamma_{sv(s)}\neq\gamma_{sv(l)}$. Because the surface and liquid both contain Ga, $\gamma_{gl}$ is assumed to be equal to $\gamma_{sl}$. Equation 1 then yields $F_{\theta}\sim r\cdot(\gamma_{gv}-\gamma_{sv})=r\cdot(\gamma_{sv(s)}-\gamma_{sv(l)})$. Tersoff {\it et al.} relate this to the difference between the Ga chemical potential of the newly-covered surface $\mu_{s}$, which is a function of $T$, and the Ga chemical potential of the newly-exposed surface $\mu_{l}$, which is taken to be the liquidus chemical potential. In particular, they use an equilibrium expansion of the chemical potential around $\mu_{l}$ to find
\begin{eqnarray}
F_{\theta}\sim r\cdot(\gamma_{sv(s)}-\gamma_{sv(l)})\\ \propto r\cdot\left(\frac{d^{2}E}{d\mu^{2}}\Big |_{\mu_{l}}-\frac{d^{2}n}{d\mu^{2}}\Big |_{\mu_{l}}\mu_{l}\right)(\mu_{s}-\mu_{l})^{2}
\end{eqnarray}
where $E$ is the Ga surface formation energy and $n$ is the excess Ga surface density. They then assume $\mu_{s}-\mu_{l}\propto T-T_{c}$ to derive their main result, namely that the driving force for droplet motion varies quadratically with temperature around $T_{c}$. The strength of this driving force is determined by the pre-factor in Equation 7. Unfortunately, the value of this pre-factor cannot be explained by any current theory. Indeed, there appears to be very little discussion in the literature investigating the variation of $E$ and $n$ with $\mu$ \cite{tersoffdiscussion}.

While this model sets important groundwork for understanding running droplets on binary semiconductors, we believe it is incomplete. In particular, it neglects two crucial experimental observations. First, the model does not incorporate any directionality in the droplet motion. Because the droplet motion is assumed to be started by thermal fluctuations, the direction of the droplet trajectory should be randomly oriented. This prediction contradicts experimental observations, in which droplets are seen to clearly move preferentially along crystallographic directions \cite{tersoff}. Second, the model does not incorporate topographic changes in the semiconductor surface. Specifically, observations of droplets moving on GaAs \cite{tersoff} and on other binary compounds \cite{steps} have shown that trenches or step structures possibly form in the surface as droplets move over it. The steps are typically oriented normal to the direction of droplet motion \cite{steps}, similar to the step structure discussed in Section 2.4 for graphene and schematically shown in Figure 1.

Taken together, we believe these observations as well as our model for graphene etching imply a possible modification to Tersoff {\it et al.}'s model. This extended model may better capture all the experimental observations for running droplets on binary semiconductor surfaces. We propose that surface reconstruction, such as the formation of trenches and steps in the GaAs surface, occurs fastest along specific crystallographic directions, as is common for many crystals. This process sets the initial direction of droplet motion -- {\it not} thermal fluctuations. Furthermore, we use the model of Section 2.4 here to determine the droplet velocity. This gives $V\sim\alpha(1-e^{-\beta/V})$ with $\alpha\equiv\Delta_{0}r/\xi$ and $\beta\equiv r/\tau$, where $\tau$ is related to the rate at which Ga is removed from a step. Unlike the graphene etching case, because the GaAs substrate acts as a reservoir for the Ga droplet, $\Delta_{0}$ is not necessarily associated with the density of dangling bonds associated with a step edge as assumed in Section 2.4. Rather, we hypothesize that this phenomenological parameter incorporates the temperature dependence of the Ga droplet velocity. For example, we suggest that $\Delta_{0}\propto(\mu_{s}-\mu_{l})^{2}$, in agreement with physical intuition: when $\mu_{s}=\mu_{l}$, Ga atoms at the GaAs step edge will not have an energetic preference to be in either the droplet or in the GaAs surface. This represents an effective GaAs edge energy $\Delta_{0}$ that is zero at $\mu_{s}$, and increases for variations of $\mu_{s}$ around $\mu_{l}$. Motivated by the discussion of Tersoff {\it et al.}, we assume that $\mu_{s}-\mu_{l}\propto T-T_{c}$. The treatment presented here thus yields a droplet velocity $V$ that increases monotonically with $(T-T_{c})^2$, consistent with the data presented in \cite{tersoff}.

\section{Conclusion}
The development of a straightforward technique to carve out straight, large-scale features with crystallographic edges in graphene would be invaluable. For example, this would represent a way of achieving graphene devices with tunable electronic band structures and exceptional performance. Recent experiments using molten metallic nanoparticles to construct graphene nanostructures with crystallographic edges are one step toward this goal, and as such, understanding the mechanism by which this happens is crucial. The simple model presented here gives a qualitative understanding of two factors driving droplet motion and etching: (1) a difference between the equilibrium wettability of graphene and the substrate that supports it and (2) the large surface energy associated with the graphene step edge. It is unclear whether both of these factors play a role in driving etching. Crucially, the time-averaged etching velocities resulting from each of these factors depend on different physical parameters. The model presented in this paper makes clear predictions for each of the factors that can be tested, and we anticipate that the results presented here will help guide experimental efforts using nanoparticles to etch graphene. Furthermore, we propose that this model is applicable to other materials systems. As an example, we apply it to the case of running droplets formed during the evaporation of binary semiconductors. We use our model to extend a current theory of this process, showing how the work presented in this paper can be used to improve the agreement of this theory with experimental observations.

\section{Acknowledgement}
It is a pleasure to acknowledge S. Mandre and F. Spaepen for useful discussions and E. A. Millman for a careful reading of the manuscript. This work was partially supported by the Peirce and Purcell fellowships of the Department of Physics at Harvard University. 
\newpage
\appendix
\section{Comparison To Previous Experiments.}
\begin{table}
  \caption{Literature values of $\theta_{g}$ and $\theta_{s}$ for droplets of different materials on a variety of substrates at high temperatures.}
  \begin{center}

    \begin{tabular}{ | c | c | c | c | l |}
    \hline
    {\bf Droplet}&{\bf Substrate}&{\bf $\theta_{g}$}&{\bf $\theta_{s}$}&{\bf Reference} \\ \hline\hline
Fe&Graphite&40-70$^\circ$&---&\cite{angle1,angle5,angle12,angle10,angle13}\\
Ni&Graphite&45-80$^\circ$&---&\cite{angle7,angle12,angle10,angle13}\\
Co&Graphite&35-80$^\circ$&---&\cite{angle10,angle12,angle13}\\
Ag&Graphite&130-170$^\circ$&---&\cite{angle6,angle11,angle12}\\
    \hline
    Fe&SiO$_{2}$&---&115-180$^\circ$&\cite{angle3,angle2}\\
Ni&SiO$_{2}$&---&110-125$^\circ$&\cite{angle3,angle8, angle11}\\
Co&SiO$_{2}$&---&100-120$^\circ$&\cite{angle3,angle4}\\
Ag&SiO$_{2}$&---&140$^\circ$&\cite{angle3}\\
\hline
Fe&SiC&---&60-70$^\circ$&\cite{angle3, angle9}\\
Ni&SiC&---&40$^\circ$&\cite{angle14}\\
Co&SiC&---&$40^\circ$&\cite{angle14}\\
Ag&SiC&---&120-140$^\circ$&\cite{angle11,angle14}\\
\hline
Fe&Al$_{2}$O$_{3}$&---&107$^\circ$&\cite{angle3}\\
Ni&Al$_{2}$O$_{3}$&---&110$^\circ$&\cite{angle3}\\
Co&Al$_{2}$O$_{3}$&---&110$^\circ$&\cite{angle3}\\
Ag&Al$_{2}$O$_{3}$&---&130$^\circ$&\cite{angle3}\\
\hline
Fe&Si$_{3}$N$_{4}$&---&90$^\circ$&\cite{angle15}\\
Ni&Si$_{3}$N$_{4}$&---&100-130$^\circ$&\cite{angle16,angle17,angle19}\\
Ag&Si$_{3}$N$_{4}$&---&130-155$^\circ$&\cite{angle18,angle14}\\
\hline
    \end{tabular}
    \end{center}

    \end{table}

To assess the applicability of the results presented in Section 2.3 -- in particular, Equation 5 -- to graphene etching, we examined the literature for contact angle measurements for nanoparticle materials relevant to experiments. These are Fe, Ni, Co, or Ag. The results are given in Table 1. Systematic contact angle measurements of molten nanoparticles on single- or few-layer graphene have, to our knowledge, not yet been performed. Thus, to get a sense of likely values of $\theta_{g}$, we turn to high-temperature measurements on bulk graphite substrates. Indeed, recent work suggests that the thickness of the graphene may not play a significant role in altering its wettability \cite{surfaceenergy1}. These data do not appear to be strongly sensitive to temperature, so we expect them to be similar to the actual contact values in graphene etching.

The data presented in Table 1 are typically for ``clean" droplets on the respective substrates. Various chemical processes might change these values. For example, significant concentrations of carbon saturated in the metal droplet are known to modify the value of $\theta_{g}$. However, values for ``clean" unsaturated droplets are probably most relevant for etching. Removal of carbon atoms via catalytic hydrogenation occurs quickly, as evidenced by observations that etched trenches extend down to the underlying substrate. Thus it is quite likely that the metal catalyst at a step edge is not saturated with an appreciable amount of carbon during the etching process.

As discussed in Section 2.3, $V>0$ only when $\theta_{g}<\theta_{s}$. The data presented in Table 1 suggest that graphene etching by Fe, Ni or Co satisfy this condition, and thus the mechanism discussed in Section 2.3 may play a role in driving the motion of these droplets. On the other hand, graphene etching by Ag does {\it not} satisfy this condition. Crucially, this implies that etching by Ag is driven solely by the mechanism presented in Section 2.4 or by some other mechanism not considered in this paper.

\newpage
\bibliographystyle{unsrt}

\begin{thebibliography}{10}

\bibitem{geim1}
A.~K. Geim.
\newblock Graphene: Status and prospects.
\newblock {\em Science}, 324:1530, 2009.

\bibitem{geim2}
A.~K. Geim and K.~S. Novoselov.
\newblock The rise of graphene.
\newblock {\em Nature Materials}, 6:183, 2007.

\bibitem{nr1}
Y.~W. Son, M.~L. Cohen, and S.~G. Louie.
\newblock Energy gaps in graphene nanoribbons.
\newblock {\em Phys. Rev. Lett.}, 97:216803, 2006.

\bibitem{nr2}
V.~Barone, O.~Hod, and G.~E. Scuseria.
\newblock Electronic structure and stability of semiconducting graphene
  nanoribbons.
\newblock {\em Nano Letters}, 6:2748, 2006.

\bibitem{nr3}
Q.~M. Yan, B.~Huang, J.~Yu, F.~W. Zheng, J.~Zang, J.~Wu, B.~L. Gu, F.~Liu, and
  W.~H. Duan.
\newblock Intrinsic current-voltage characteristics of graphene nanoribbon
  transistors and effect of edge doping.
\newblock {\em Nano Letters}, 7:1469, 2007.

\bibitem{nr4}
Y.~W. Son, M.~L. Cohen, and S.~G. Louie.
\newblock Half-metallic graphene nanoribbons.
\newblock {\em Nature}, 444:347, 2006.

\bibitem{nr5}
M.~Y. Han, J.~C. Brant, and P.~Kim.
\newblock Electron transport in disordered graphene nanoribbons.
\newblock {\em Phys. Rev. Lett.}, 104:056801, 2010.

\bibitem{nr6}
D.~Gunlycke, D.~A. Areshkin, J.~Li, J.~W. Mintmire, and C.~T. White.
\newblock Graphene nanostrip digital memory device.
\newblock {\em Nano Letters}, 7:3608, 2007.

\bibitem{nr7}
D.~Basu, M.~J. Gilbert, L.~F. Register, S.~K. Banerjee, and A.~H. MacDonald.
\newblock Effect of edge roughness on electronic transport in graphene
  nanoribbon channel metal-oxide-semiconductor field-effect transistors.
\newblock {\em Appl. Phys. Lett.}, 92:042114, 2008.

\bibitem{nr8}
T.~C. Li and S.~P. Lu.
\newblock Quantum conductance of graphene nanoribbons with edge defects.
\newblock {\em Phys. Rev. B}, 77:085408, 2008.

\bibitem{nr9}
K.~Nakada, M.~Fujita, G.~Dresselhaus, and M.~S. Dresselhaus.
\newblock Edge state in graphene ribbons: Nanometer size effect and edge shape
  dependence.
\newblock {\em Phys. Rev. B}, 54:17954, 1996.

\bibitem{nr10}
L.~Brey and H.~A. Fertig.
\newblock Electronic states of graphene nanoribbons studied with the {D}irac
  equation.
\newblock {\em Phys. Rev. B}, 73:235411, 2006.

\bibitem{datta1}
S.~S. Datta, D.~R. Strachan, S.~M. Khamis, and A.~T.~C. Johnson.
\newblock Crystallographic etching of few-layer graphene.
\newblock {\em Nano Letters}, 8:1912, 2008.

\bibitem{previous1}
L.~Ci, Z.~Xu, L.~Wang, W.~Gao, F.~Ding, K.~F. Kelly, B.~I. Yakobson, and P.~M.
  Ajayan.
\newblock Controlled nanocutting of graphene.
\newblock {\em Nano Res.}, 1:116, 2008.

\bibitem{degennes1}
P.~G. de~Gennes.
\newblock The dynamics of reactive wetting on solid surfaces.
\newblock {\em Physica A}, 249:196, 1998.

\bibitem{ondarcuhu}
F.~D.~Dos Santos and T.~Ondarcuhu.
\newblock Free-running droplets.
\newblock {\em Phys. Rev. Lett.}, 75:2972, 1995.

\bibitem{klinke}
C.~Klinke, J.~M. Bonard, and K.~Kern.
\newblock Formation of metallic nanocrystals from gel-like precursor films for
  {C}{V}{D} nanotube growth: An in situ {T}{E}{M} characterization.
\newblock {\em J. Phys. Chem. B}, 108:11357, 2004.

\bibitem{molten1}
Y.~Shibuta and T.~Suzuki.
\newblock Melting and nucleation of iron nanoparticles: A molecular dynamics
  study.
\newblock {\em Chem. Phys. Lett.}, 445:265, 2007.

\bibitem{molten2}
Y.~Homma, Y.~Kobayashi, T.~Ogino, D.~Takagi, R.~Ito, Y.~J. Jung, and P.~M.
  Ajayan.
\newblock Role of transition metal catalysts in single-walled carbon nanotube
  growth in chemical vapor deposition.
\newblock {\em J. Phys. Chem. B}, 107:12161, 2003.

\bibitem{reactive1}
L.~R. Radovic and B.~Bockrath.
\newblock On the chemical nature of graphene edges: Origin of stability and
  potential for magnetism in carbon materials.
\newblock {\em JACS}, 127:5917, 2005.

\bibitem{reactive2}
D.~E. Jiang, B.~G. Sumpter, and S.~Dai.
\newblock Unique chemical reactivity of a graphene nanoribbon's zigzag edge.
\newblock {\em J. Chem. Phys.}, 126:134701, 2007.

\bibitem{reactive3}
R.~Sharma, N.~Nair, and M.~S. Strano.
\newblock Structure-reactivity relationships for graphene nanoribbons.
\newblock {\em J. Phys. Chem. C}, 113:14771, 2009.

\bibitem{jarillo}
L.~C. Campos, V.~R. Manfrinato, J.~D. Sanchez-Yamagishi, J.~Kong, and
  P.~Jarillo-Herrero.
\newblock Anisotropic etching and nanoribbon formation in single-layer
  graphene.
\newblock {\em Nano Letters}, 9:2600, 2009.

\bibitem{previous2}
A.~L. Elias, A.~R. Botello-Mendez, D.~Meneses-Rodriguez, V.~J. Gonzalez,
  D.~Ramirez-Gonzalez, L.~Ci, E.~Munoz-Sandoval, P.~M. Ajayan, H.~Terrones, and
  M.~Terrones.
\newblock Longitudinal cutting of pure and doped carbon nanotubes to form
  graphitic nanoribbons using metal clusters as nanoscalpels.
\newblock {\em Nano Letters}, 10:366, 2010.

\bibitem{previous3}
F.~Schaffel, J.~H. Warner, A.~Bachmatiuk, B.~Rellinghaus, B.~Buchner,
  L.~Schultz, and M.~H. Rummeli.
\newblock Shedding light on the crystallographic etching of multi-layer
  graphene at the atomic scale.
\newblock {\em Nano Res.}, 2:695, 2009.

\bibitem{previous4}
F.~Schaffel, J.~H. Warner, A.~Bachmatiuk, B.~Rellinghaus, B.~Buchner,
  L.~Schultz, and M.~H. Rummeli.
\newblock On the catalytic hydrogenation of graphite for graphene nanoribbon
  fabrication on the catalytic hydrogenation of graphite for graphene
  nanoribbon fabrication.
\newblock {\em physica status solidi (b)}, 246:2540, 2009.

\bibitem{previous5}
L.~Ci, L.~Song, D.~Jariwala, A.~L. Elias, W.~Gao, M.~Terrones, and P.~M.
  Ajayan.
\newblock Graphene shape control by multistage cutting and transfer.
\newblock {\em Adv. Mater.}, 21:4487, 2009.

\bibitem{previous6}
N.~Severin, S.~Kirstein, I.~M. Sokolov, and J.~P. Rabe.
\newblock Rapid trench channeling of graphenes with catalytic silver
  nanoparticles.
\newblock {\em Nano Letters}, 9:457, 2009.

\bibitem{previous7}
J.~H. Warner, M.~H. Rummeli, A.~Bachmatiuk, M.~Wilson, and B.~Buchner.
\newblock Examining {C}o-based nanocrystals on graphene using low-voltage
  abberation-corrected transmission electron microscopy.
\newblock {\em ACS Nano}, 4:470, 2010.

\bibitem{meltingdepression}
P.~Buffat and J.~P. Borel.
\newblock Size effect on the melting temperature of gold particles.
\newblock {\em Phys. Rev. A}, 13:2287, 1976.

\bibitem{anisotropic1}
J.~M. Thomas.
\newblock {\em Chemistry and Physics of Carbon}, volume~1.
\newblock M. Dekker, New York, 1965.

\bibitem{density}
D.~R. Lide and W.~M. Haynes, editors.
\newblock {\em CRC Handbook of Chemistry and Physics}.
\newblock CRC Press, Boca Raton, 90th edition, 2009-2010.

\bibitem{angle2}
Y.~Wang, B.~Li, P.~S. Ho, Z.~Yao, and L.~Shi.
\newblock Effect of supporting layer on growth of carbon nanotubes by thermal
  chemical vapor deposition.
\newblock {\em Appl. Phys. Lett.}, 89:183113, 2006.

\bibitem{mele}
Z.~Luo, L.~A. Somers, Y.~Dan, T.~Ly, N.~J. Kybert, E.~J. Mele, and A.~T.~C.
  Johnson.
\newblock Size-selective nanoparticle growth on few-layer graphene films.
\newblock {\em Nano Letters}, 10:777, 2010.

\bibitem{dissipation1}
P.~G. de~Gennes.
\newblock Wetting: statics and dynamics.
\newblock {\em Rev. Mod. Phys.}, 57:827, 1985.

\bibitem{dissipation2}
C.~Huh and L.~E. Scriven.
\newblock Hydrodynamic model of steady movement of a solid/liquid/fluid contact
  line.
\newblock {\em J. Colloid Interface Sci.}, 35:85, 1971.

\bibitem{dissipation3}
F.~Brochard-Wyart and P.~G. de~Gennes.
\newblock Spreading of a drop between a solid and a viscous polymer.
\newblock {\em Langmuir}, 10:2440, 1994.

\bibitem{sutherland}
W.~Sutherland.
\newblock The viscosity of mixed gases.
\newblock {\em Philos. Mag.}, 5:507, 1895.

\bibitem{edgeenergy1}
P.~Koskinen, S.~Malola, and H.~Hakkinen.
\newblock Self-passivating edge reconstructions of graphene.
\newblock {\em Phys. Rev. Lett.}, 101:115502, 2008.

\bibitem{edgeenergy2}
T.~Kawai, Y.~Miyamoto, O.~Sugino, and Y.~Koga.
\newblock Graphitic ribbons without hydrogen-termination: Electronic structures
  and stabilities.
\newblock {\em Phys. Rev. B}, 62:R16349, 2000.

\bibitem{edgeenergy3}
C.~K. Gan and D.~J. Srolovitz.
\newblock First-principles study of graphene edge properties and flake shapes.
\newblock {\em Phys. Rev. B}, 81:125445 (2010).

\bibitem{edgeenergy4}
S.~V. Rotkin.
\newblock On surface energy of graphene and carbon nanoclusters.
\newblock {\em arxiv:cond-mat/0107312}, 2001.

\bibitem{edgeenergy5}
V.~L. Kuznetsov, A.~N. Usoltseva, A.~L. Chuvilin, E.~D. Obraztsova, and J.~M.
  Bonard.
\newblock Thermodynamic analysis of nucleation of carbon deposits on metal
  particles and its implications for the growth of carbon nanotubes.
\newblock {\em Phys. Rev. B}, 64:235401, 2001.

\bibitem{realtime}
D.~R. Strachan, D.~E. Johnston, B.~S. Guiton, S.~S. Datta, P.~K. Davies, D.~A.
  Bonnell, and A.~T.~C. Johnson.
\newblock Real-time {TEM} imaging of the formation of crystalline nanoscale
  gaps.
\newblock {\em Phys. Rev. Lett.}, 100:056805, 2008.

\bibitem{al2o3}
L.~Liao, J.~Bai, Y.~Qu, Y.~Huang, and X.~Duan.
\newblock Single-layer graphene on {A}l$_{2}${O}$_{3}$/{S}i substrate: better
  contrast and higher performance of graphene transistors.
\newblock {\em Nanotechnology}, 21:015705, 2010.

\bibitem{si3n4}
Alexander Klekachev, Mirco Cantoro, Amirhasan Nourbakhsh, Marleen H.~Van der
  Veen, Francesca Clemente, Andre~L. Stesmans, Bert Sels, Marc Heyns, and
  Stefan~De Gendt.
\newblock Substrate-limited mobility in single- and bi-layer graphene on
  dielectric materials.
\newblock {\em ECS Trans.}, 19(5):201, 2009.

\bibitem{deheer}
C.~Berger, Z.~Song, X.~Li, X.~Wu, N.~Brown, C.~Naud, D.~Mayou, T.~Li, J.~Hass,
  A.~N. Marchenkov, E.~H. Conrad, P.~N. First, and W.~A. de~Heer.
\newblock Electronic confinement and coherence in patterned epitaxial graphene.
\newblock {\em Science}, 312:1191, 2006.

\bibitem{surfaceenergy1}
Y.~J. Shin, Y.~Wang, H.~Huang, G.~Kalon, A.~T.~S. Wee, Z.~Shen, C.~S. Bhatia,
  and H.~Yang.
\newblock Surface-energy engineering of graphene.
\newblock {\em Langmuir}, 26:3798, 2010.

\bibitem{surfaceenergy2}
J.~Rafiee, M.~A. Rafiee, Z.~Z. Yu, and N.~Koratkar.
\newblock Superhydrophobic to superhydrophilic wetting control in graphene
  films.
\newblock {\em Adv. Mater.}, 22:2151, 2010.

\bibitem{tersoff}
J.~Tersoff, D.~E. Jesson, and W.~X. Tang.
\newblock Running droplets of gallium from evaporation of gallium arsenide.
\newblock {\em Science}, 324:236, 2009.

\bibitem{steps}
E.~Hilner, A.~A. Zakharov, K.~Schulte, P.~Kratzer, J.~N. Anderson, E.~Lundgren,
  and A.~Mikkelsen.
\newblock Ordering of the nanoscale step morphology as a mechanism for droplet
  self-propulsion.
\newblock {\em Nano Letters}, 9:2710, 2009.

\bibitem{tersoffdiscussion}
J.~Tersoff.
\newblock Private communication.
\newblock March 11, 2010.

\bibitem{angle1}
L.~Zhao and V.~Sahajwalla.
\newblock Interfacial phenomena during wetting of graphite/alumina mixtures by
  liquid iron.
\newblock {\em ISIJ International}, 43:1, 2003.

\bibitem{angle5}
V.~I. Nizhenko and L.~I. Floka.
\newblock Contact reaction of graphite with liquid iron and iron base melts.
\newblock {\em Powder Metallurgy and Metal Ceramics}, 13:1068, 1974.

\bibitem{angle12}
E.~R. Parker and U.~Colombo, editors.
\newblock {\em The science of materials used in advanced technology}.
\newblock Wiley (New York), 1973.

\bibitem{angle10}
V.~I. Kostikov, M.~A. Maurakh, and A.~V. Nozhkina.
\newblock Wetting of diamond and graphite by liquid iron-titanium alloys.
\newblock {\em Powder Metallurgy and Metal Ceramics}, 10:1573, 1971.

\bibitem{angle13}
J.~R. Tinklepaugh and W.~B. Crandall, editors.
\newblock {\em Cermets}.
\newblock Reinhold (New York), 1960.

\bibitem{angle7}
Y.~V. Naidich, V.~M. Perevertailo, and G.~M. Nevodnik.
\newblock Kinetics of spreading of molten metals on solid surfaces.
\newblock {\em Powder Metallurgy and Metal Ceramics}, 11:555, 1972.

\bibitem{angle6}
C.~P. Buhsmer and E.~A. Heintz.
\newblock Application of wetting theory to the non-reactive liquid
  metal-graphite systems.
\newblock {\em J. Mater. Sci.}, 4:592, 1969.

\bibitem{angle11}
R.~Asthana.
\newblock An empirical correlation between contact angles and surface tension
  in some ceramic-metal systems.
\newblock {\em Metallurgical and Materials Transactions A}, 25:225, 1994.

\bibitem{angle3}
N.~Eustathopoulos and B.~Drevet.
\newblock Interfacial bonding, wettability and reactivity in metal/oxide
  systems.
\newblock {\em Journal de Physique III}, 4:1865, 1994.

\bibitem{angle8}
S.~J. Hitchcock, N.~T. Carroll, and M.~G. Nicholas.
\newblock Some effects of substrate roughness on wettability.
\newblock {\em J. Mater. Sci.}, 16:714, 1981.

\bibitem{angle4}
C.~Favazza, J.~Trice, H.~Krishna, R.~Kalyanaraman, and R.~Sureshkumar.
\newblock Laser-induced short- and long-range orderings of {C}o nanoparticles
  on {S}i{O}$_{2}$.
\newblock {\em Appl. Phys. Lett.}, 88:153118, 2006.

\bibitem{angle9}
S.~Kalogeropoulou, L.~Baud, and N.~Eustathopoulos.
\newblock Relationship between wettability and reactivity in {Fe}/{S}i{C}
  system.
\newblock {\em Acta metall. mater.}, 43:907, 1995.

\bibitem{angle14}
N.~Eustathopoulos, M.~G. Nicholas, and B.~Drevet.
\newblock {\em Wettability at High Temperatures}.
\newblock Pergamon (Oxford), 1999.

\bibitem{angle15}
G.~A. Yasinskaya.
\newblock The wetting of refractory carbides, borides, and nitride by molten
  metals.
\newblock {\em Powder Metallurgy and Metal Ceramics}, 5:557, 1966.

\bibitem{angle16}
Y.~V. Naidich, V.~S. Zhuravlev, N.~I. Frumina, B.~D. Kostyuk, N.~A.
  Krasovskaya, and V.~G. Ostrovskii.
\newblock Wetting of ceramics based on silicon nitride by metallic melts.
\newblock {\em Powder Metallurgy and Metal Ceramics}, 27:888, 1988.

\bibitem{angle17}
P.~P. Pikuza, A.~D. Panasyuk, and I.~P. Neshpor.
\newblock Contact interaction between materials based on silicon nitride and
  sianols with {A}l, {S}i, and {N}i.
\newblock {\em Refractories and Industrial Ceramics}, 26:189, 1985.

\bibitem{angle19}
G.~V. Samsonov, A.~D. Panasyuk, and I.~M. Finkel'shtein.
\newblock Adhesion of liquid nickel to materials of the
  {S}i$_{3}${N}$_{4}$-{Z}r{O}$_{2}$ and {S}i$_{3}${N}$_{4}$-{C}r$_{2}${O}$_{3}$
  systems.
\newblock {\em Powder Metallurgy and Metal Ceramics}, 16:392, 1977.

\bibitem{angle18}
M.~Naka, M.~Kubo, and I.~Okamoto.
\newblock Wettability of silicon nitride by aluminum, copper and silver.
\newblock {\em J. Mater. Sci. Lett.}, 6:965, 1987.

\end{thebibliography}

\newpage
\begin{figure}
\begin{center}
\includegraphics[width=4in]{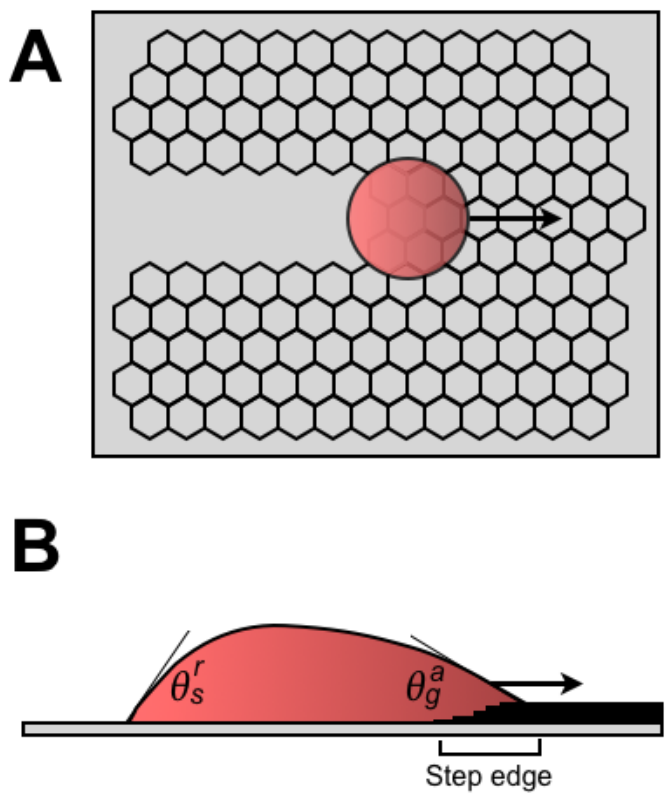}
\caption{{\bf Schematic of the experiment.} (A) Top-down view of graphene etching by a molten nanoparticle, showing possible step structure of graphene layers at the edge undergoing etching. (B) Two-dimensional representation of etching, showing advancing and receding contact angles. The grey underlying layer is the substrate while the black line being etched away is the graphene, showing an example of a step edge. These diagrams are not to scale.}
\end{center}
\end{figure}
\newpage

\begin{figure}
\begin{center}
\includegraphics[width=6in]{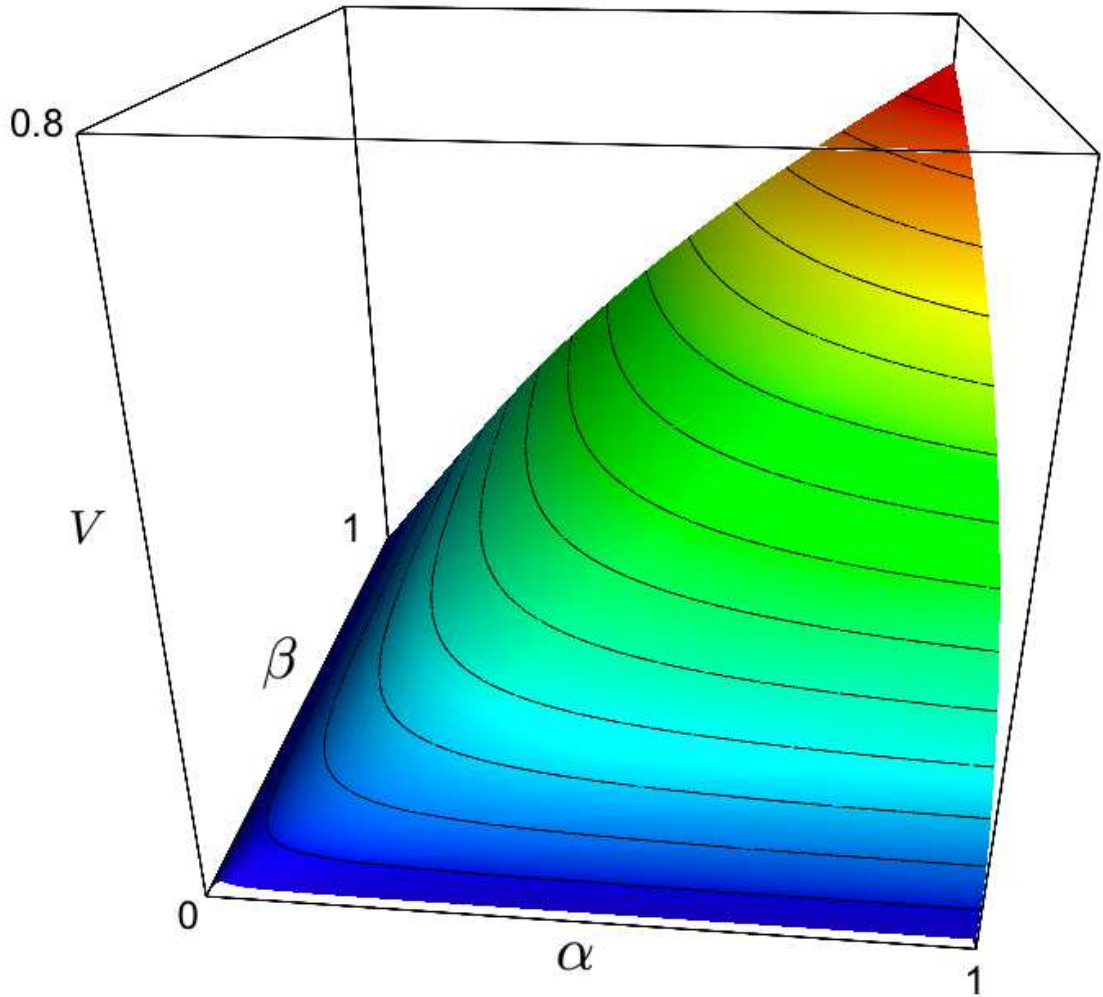}
\caption{{\bf Droplet velocity.} Dependence of the droplet etching velocity $V$ as a function of the characteristic velocities $\alpha$ and $\beta$, as given in Equation 4 of the text.}
\end{center}
\end{figure}
\newpage

\begin{figure}
\begin{center}
\includegraphics[width=6in]{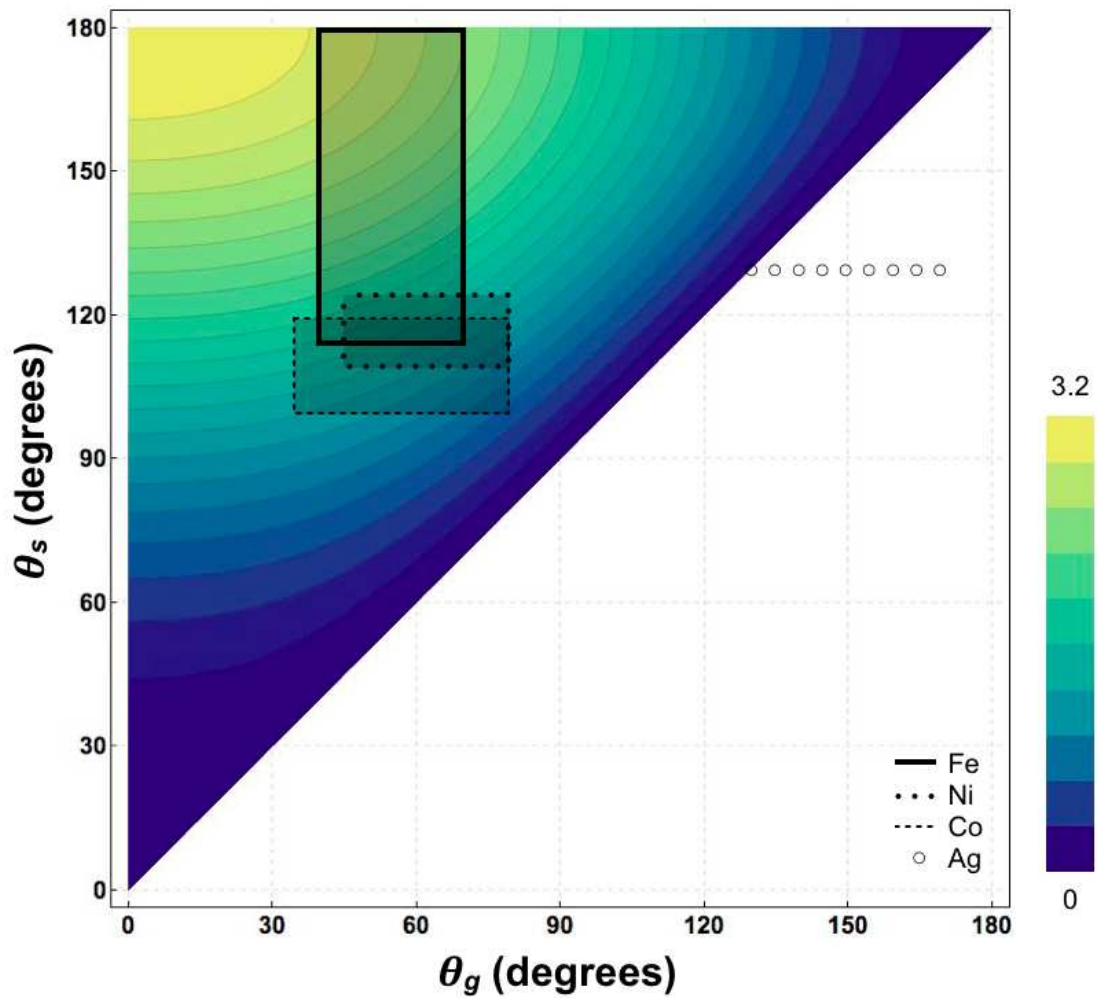}
\caption{{\bf Influence of substrate and graphene wettability.} Wettability ``state diagram", showing droplet etching velocity $V$ as given in Equation 5 of the text. Shaded region represents positive values of $V$, as given by the color scale and contours. Rectangular regions and unfilled circles indicate where different nanoparticle materials lie on this diagram.}
\end{center}
\end{figure}
\newpage

	\end{document}